\documentclass[iop]{emulateapj}       

\newcommand{\Lsol}{\mbox{$L_\odot$}}

\newcommand{\asec}{\mbox{$''$}}
\newcommand{\kms}{\mbox{km s$^{-1}$}}
\newcommand{\perkms}{\mbox{(km s$^{-1}$)$^{-1}$}}
\newcommand{\persquarecm}{\mbox{cm$^{-2}$}}
\newcommand{\percubiccm}{\mbox{cm$^{-3}$}}

\newcommand{\hr}{\mbox{$^{\rm h}$}}
\newcommand{\mn}{\mbox{$^{\rm m}$}}
\newcommand{\s}{\mbox{$^{\rm s}$}}
\newcommand{\persec}{\mbox{s$^{-1}$}}
\newcommand{\cubiccm}{\mbox{cm$^{3}$}}
\newcommand{\about}{\mbox{$\sim$}}
\newcommand{\plus}{\mbox{$+$}}
\newcommand{\minus}{\mbox{$-$}}

\newcommand{\frest}{\mbox{$f_{\rm rest}$}}   
\newcommand{\twelveC}{\mbox{$^{12}$C}}    
\newcommand{\thirteenC}{\mbox{$^{13}$C}}   
\newcommand{\fourteenN}{\mbox{$^{14}$N}}   
\newcommand{\fifteenN}{\mbox{$^{15}$N}}   
\newcommand{\HH}{\mbox{H$_2$}}
\newcommand{\HthirteenCN}{\mbox{H$^{13}$CN}}          
\newcommand{\HCfifteenN}{\mbox{HC$^{15}$N}}             
\newcommand{\HCthreeN}{\mbox{HC$_{3}$N}}                 
\newcommand{\HCOplus}{\mbox{HCO$^{+}$}}                   
\newcommand{\NtwoHplus}{\mbox{N$_{2}$H$^{+}$}}       
\newcommand{\vJfourtothree}{\mbox{($v_2$=$1^{\rm 1f}$,J=4--3)}}   
\newcommand{\vJthreetotwo}{\mbox{($v_2$=$1^{\rm 1f}$,J=3--2)}}    
\newcommand{\vtwo}{\mbox{$v_2$}}      
\newcommand{\vtwol}{\mbox{$v_2^{\;\;l}$}}      
\newcommand{\el}{\mbox{$l$}}      
\newcommand{\Avib}{\mbox{$A_{\rm vib}$}}  
\newcommand{\Arot}{\mbox{$A_{\rm rot}$}}  
\newcommand{\Tex}{\mbox{$T_{\rm ex}$}}  
\newcommand{\Trot}{\mbox{$T_{\rm rot}$}}  
\newcommand{\Tvib}{\mbox{$T_{\rm vib}$}}  
\newcommand{\Tkin}{\mbox{$T_{\rm kin}$}}  
\newcommand{\nHH}{\mbox{$n_{\rm H_2}$}}  
\newcommand{\nd}{\nodata}

\newcommand{\Tb}{\mbox{$T_{\rm b}$}}

\shorttitle{Vibrationally Excited HCN in NGC 4418}
\shortauthors{SAKAMOTO et al.}

\slugcomment{Accepted for publication in {\it ApJ Letters}}

\begin{document}
\title{Vibrationally Excited HCN in the Luminous Infrared Galaxy NGC 4418}

\author{Kazushi Sakamoto\altaffilmark{1}, 
Susanne Aalto\altaffilmark{2}, 
Aaron S. Evans\altaffilmark{3,4},
Martina C. Wiedner\altaffilmark{5}, 
and
David J. Wilner\altaffilmark{6}
}
\altaffiltext{1}{Academia Sinica, Institute of Astronomy and Astrophysics, Taipei, Taiwan} 
\altaffiltext{2}{Onsala Space Observatory, Onsala, Sweden} 
\altaffiltext{3}{Department of Astronomy, University of Virginia, Charlottesville, VA, USA}
\altaffiltext{4}{National Radio Astronomy Observatory, Charlottesville, VA, USA}
\altaffiltext{5}{Observatoire de Paris, Paris, France}
\altaffiltext{6}{Harvard-Smithsonian Center for Astrophysics, Cambridge, MA, USA}

\begin{abstract} 
Infrared pumping and its effect on the excitation of HCN molecules can be
important when using rotational lines of HCN to probe dense molecular gas
in galaxy nuclei.
We report the first extragalactic detection of (sub)millimeter rotational lines of vibrationally 
excited HCN, in the dust-enshrouded nucleus of the luminous infrared galaxy NGC 4418. 
We estimate the excitation temperature of 
$\Tvib \approx 230$ K between the
vibrational ground  and excited ($v_2$=1) states. 
This excitation is most likely due to infrared radiation.
At this high vibrational temperature the path through the  $v_2$=1 state must
have a strong impact on the rotational excitation in the vibrational ground level, 
although it may not be dominant for all rotational levels.
Our observations also revealed nearly confusion limited lines of
CO, HCN, \HCOplus, \HthirteenCN, \HCfifteenN, CS, \NtwoHplus, and \HCthreeN\
at $\lambda \sim 1$ mm.
Their relative intensities may also be affected by the infrared pumping.
\end{abstract}


\section{Introduction}  
\label{s.introduction}
\begin{figure}[t]
\begin{center}
\epsscale{0.6}
\plotone{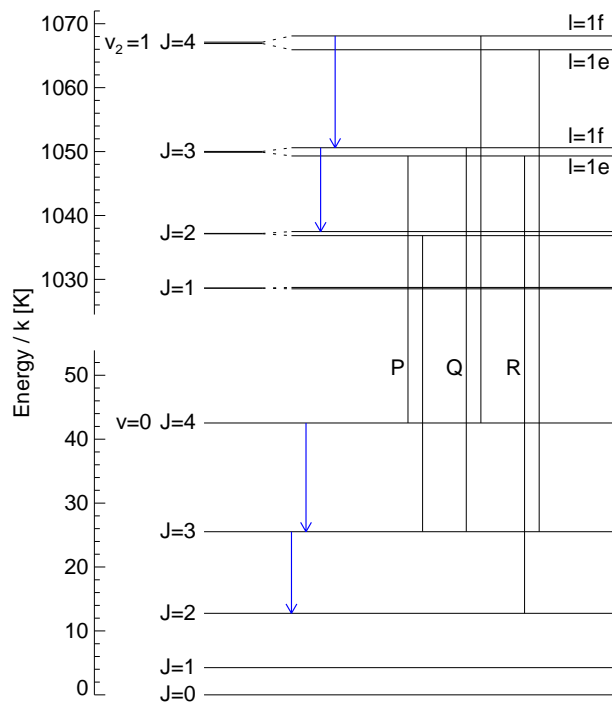}  
\end{center}
\caption{ \label{f.hcnLevels}
HCN energy level diagram, with $l$-doublets magnified 10 times on the right.
Blue arrows are the rotational transitions that we detected.
Allowed vibrational transitions  are 
$\vtwo=0 \leftrightarrow 1^{1f},\; \Delta J=0$
(Q branch)
and 
$\vtwo=0 \leftrightarrow 1^{1e},\; \Delta J=\pm 1$
(P and R branches); some of them are plotted.
}
\end{figure}

Hydrogen cyanide (HCN) is a major tracer of dense molecular gas in space.
It decays fast in its rotational energy ladder,
300 times faster than carbon monoxide\footnote{Molecular parameters in this Letter
are from the Cologne Database of Molecular Spectroscopy \citep{CDMS05}
through
Splatalogue ({\tt http://www.splatalogue.net})
unless otherwise noted.},
because of its large dipole moment.
Rotationally excited HCN therefore suggests
an excitation mechanism fast enough to counter the decay,
such as frequent collisions with interstellar \HH\ in dense molecular gas.
HCN or any molecule with a large dipole moment, however, will not trace dense gas if there 
is another excitation mechanism that is faster than the \HH\ collisions and independent of gas density.
One such excitation path is through a vibrationally excited state, 
to which molecules can be pumped by infrared radiation
\citep{Carroll81}.
This excitation is possible because a round trip of radiative transitions between 
the vibrational ground and excited states can change the rotational level J by 2.
The first vibrationally excited state of HCN is its bending state (\vtwo=1) 1024 K above the ground
(Fig. \ref{f.hcnLevels}).
Radiative excitation of HCN through this state has been 
observed around dusty evolved and young high-mass stars \citep{Ziurys86}.
Meanwhile, despite the increasing number of extragalactic HCN observations,
it has been controversial whether such radiative excitation is significant in extragalactic 
observations where the volume probed is much larger and having enough IR flux there more difficult
\citep{Aalto95,Gao04,Garcia-Burillo2006,Krips08, Gracia-Carpio08}.
It was found significant in a $10^{14}$ \Lsol\ quasar \citep{Weiss07,Riechers10} but 
the quasar may be exceptional.

In this Letter, we report our detection of rotational transitions from vibrationally excited HCN in 
the luminous infrared galaxy NGC 4418. 
We analyze HCN excitation by measuring vibrational and rotational temperatures from the data.
Our detection follows the detection of vibrationally excited \HCthreeN\ in the same galaxy by \citet{CA10}.
Related HCN lines are the \vtwo=1$\leftarrow$0 absorption at 14 \micron\
in NGC 4418 and other galaxies \citep{Lahuis07}
and 
the direct $l$-type transitions of vibrationally excited HCN in Arp 220 \citep{Salter08}.
NGC 4418 has a high luminosity and a compact and heavily dust shrouded nucleus 
\citep[$L_{\rm IR} = 10^{11.1} \Lsol$, ${\rm size} \lesssim 100$ pc =0\farcs6 ;][]{Dudley97,Evans03}.
Both an active galactic nucleus and a very young starburst have been proposed for the luminosity source
\citep{Spoon01,Roussel03}.

\section{SMA Observations} 
\label{s.obs}
We observed the nucleus of NGC 4418
(R.A. = 12\hr 26\mn 54.62\s, Dec.=\minus00\arcdeg 52\arcmin 39.4\arcsec; J2000)
using the Submillimeter Array (SMA)\footnote{
The Submillimeter Array is a joint
project between the Smithsonian Astrophysical Observatory and the
Academia Sinica Institute of Astronomy and Astrophysics, and is
funded by the Smithsonian Institution and the Academia Sinica.
} in 2010 March and May 
in the subcompact and compact configurations, respectively.
Receivers were tuned to around 350 GHz in March and 270 GHz in May 
for J=4--3 and 3--2 transitions of HCN and \HCOplus\ 
as well as CO(3--2).
Signals were separately recorded from upper and lower sidebands (USB and LSB),
whose central frequencies are 12 GHz apart from each other.
Each sideband is 4 GHz wide; we filled its central 30 MHz (\about30 \kms)  gap by interpolation.
Spectral resolution was 1.625 and 3.25 MHz for our 270 and 350 GHz runs,
respectively. 
The system gain, passband, and flux scale were calibrated using a nearby quasar 3C273 (which dimmed in May), 
quasars and planets, and Mars ($T_{\rm b \; 270\, GHz}$=190 K and $T_{\rm b \; 350\, GHz}$=192 K) 
and Titan ($T_{\rm b \; 270\, GHz}$=77 K for continuum), respectively.

We reduced the data using MIR and MIRIAD packages and 
made images with the natural weighting of visibilities 
at 30 \kms\ velocity resolution.  
The full width at half maximum of the synthesized beam was
5\farcs2 for the 350 GHz data, which had projected baselines 
of 6--25 m, and 2\farcs6 for the 270 GHz data (15--68 m).
We did not subtract continuum in order to later determine the continuum and lines in the spectra.
There are two calibration notes.
First, we took the spectral index of 3C273 into account in our gain calibration
in order to accurately scale the flux of our two sidebands.
We used the spectral index $\alpha = -1.0$ (for $S_\nu \propto \nu^\alpha$), the mean
of our March and May measurements.
Second, we verified our passband calibration by setting aside a fraction of 3C273 data 
when determining passband shapes 
and by applying the same calibration to the galaxy and the 3C273 data.
The flatness of the resulting test 3C273 spectra confirmed our calibration.

\section{Results} \label{s.results}
Strong continuum and line emission were detected at the nucleus of NGC 4418.
No extended emission was detected off the nucleus except CO(3--2) that
was marginally resolved. 
Any missing flux in our data is estimated to be less than our calibration error on the basis of 
the ratio of our flux to single-dish flux, 
$0.82\pm0.19$ for CO(3--2) \citep{Yao03} and $1.21\pm0.19$ for \HCOplus(4--3)
(Aalto et al. in prep.) 

\begin{figure}[t]
\begin{center} 
\epsscale{1.0} 
\plottwo{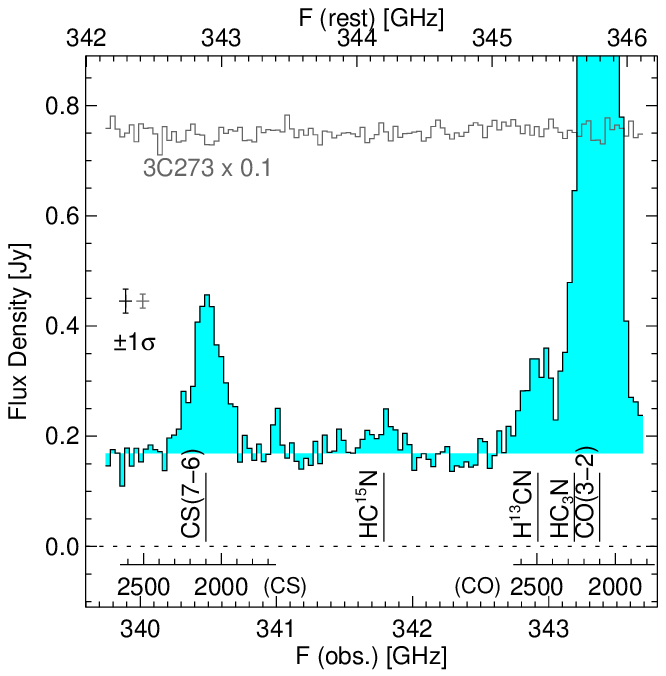}{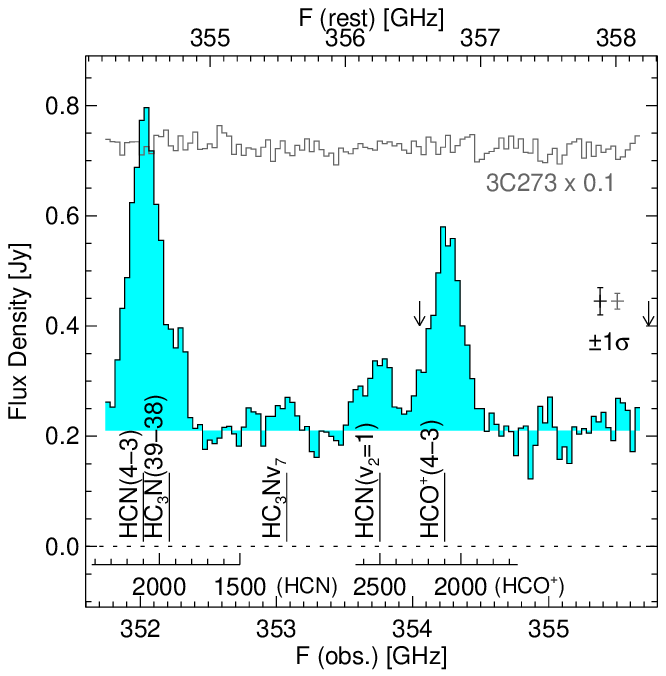}
\\
\plottwo{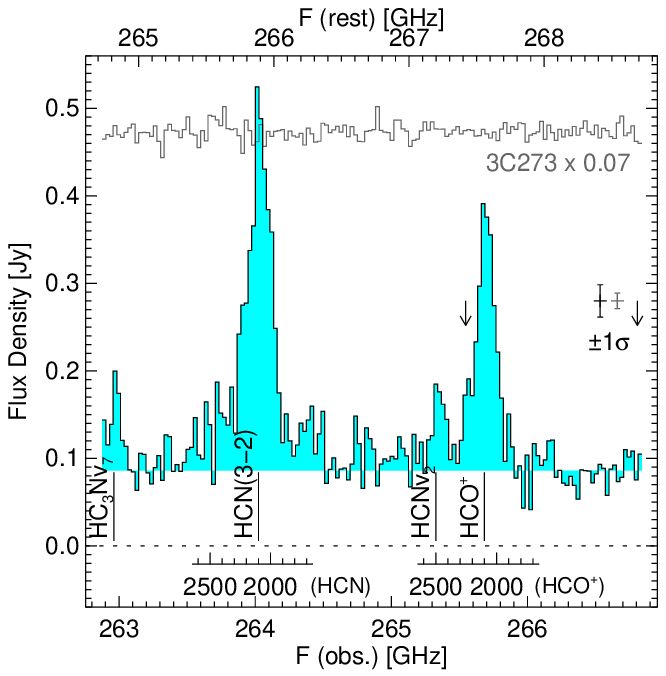}{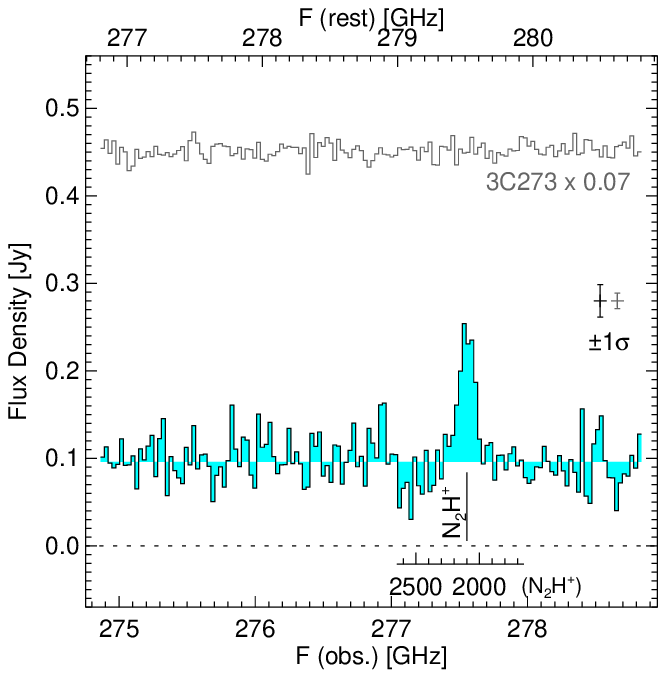}
\end{center}
\caption{ \label{f.cmpsub_spec}
Spectra of the central 6\arcsec\ (1 kpc) of NGC 4418.
Major lines are marked at the redshift of the galaxy (2100 \kms).
Full line names are in Table \ref{t.param}.
Velocity axes are given for strong lines. 
Baselines of the shaded parts of the spectra are the continuum levels.
Pairs of arrows in the top-right and bottom-left panels mark the frequencies of the \HCOplus($v_2$=1) doublet. 
Scaled spectra of 3C273 from the same observations are also plotted to verify passband calibration.
}
\end{figure}

\begin{figure}[t]
\begin{center} 
\epsscale{0.5}
\plotone{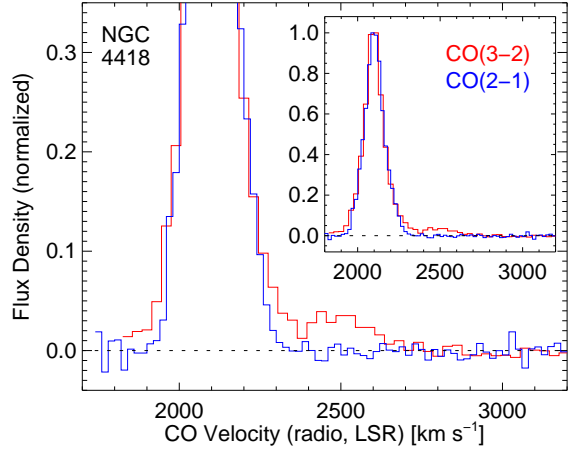} 
\end{center}
\caption{ \label{f.co32co21}
Comparison of CO(3--2) and CO(2--1) spectra, showing that CO(2--1)
has no emission (i.e., no high-velocity gas) at 2500 (=$V_{\rm sys}+400$) \kms.
The continuum-subtracted spectra of CO(3--2) and (2--1) are from the data in Fig. \ref{f.cmpsub_spec}
and 3\arcsec\ resolution SMA observations \citep{Espada10}, respectively.
}
\end{figure}

Our spectra and emission parameters in the central kpc are presented in
Fig. \ref{f.cmpsub_spec} and Table \ref{t.param}.
The strongest line is CO(3--2) followed by HCN and \HCOplus\ in both (4--3) and (3--2),
CS(7--6), and \NtwoHplus(3--2). 
We also identified \HthirteenCN\ and vibrationally excited HCN next to the CO and \HCOplus\
lines, respectively.
Although both are at about \plus400 \kms\ from the systemic velocity of the adjacent brighter lines,
they cannot be a high-velocity gas because such a feature is absent in CO(2--1) (Fig. \ref{f.co32co21}).
This HCN is excited in its bending mode (\vtwo=1) and the line
is one of the \el-doublet ($l$=1f).
The other line ($l$=1e) is only \minus38 \kms\  from the ground state HCN line and indistinguishable.
At even lower levels, our data show line features that we attribute to \HCfifteenN\ and \HCthreeN.
The data plausibly contain  more lines, e.g., on both sides of HCN(3--2).
Line density is high, and we are approaching the confusion limit.

The continuum, which we determined from `line free' channels (details are in Table \ref{t.param}),
has a  spectral index of $\alpha=2.7 \pm 0.5$ in our data excluding the 350 GHz  USB.
It is consistent with previous submillimeter measurements \citep[$\alpha \sim 2.9$;][]{Dunne00,Yang07}.
The 350 USB continuum is an outlier with $\alpha=6.25\pm0.84$ between the 350 GHz sidebands 
(cf. $\alpha=2.40\pm1.04$ between our 270 GHz sidebands).
The 350 USB `line free' channels may thus contain blended lines.
The contamination would be 23 mJy (i.e., 12\%) if we scale the LSB 
continuum with $\alpha=3$.
This possible error is not included in Table \ref{t.param} and it would affect weak lines more.

\section{HCN Excitation}
We obtained the following HCN excitation temperatures
among (\vtwol,J)=($1^{1f}$,4), ($1^{1f}$,3), and (0,4) states
under the assumptions described below:
\Tex($1^{1f}$,4; $1^{1f}$,3)=31$\pm$13 K, 
\Tex($1^{1f}$,4; 0,4)=227$\pm$18 K, 
and 
\Tex($1^{1f}$,3; 0,4)=254$\pm$24 K.
The first one is a rotational excitation temperature \Trot\  
and the second is a vibrational excitation temperature \Tvib.  
We needed opacity correction to the HCN(4--3) flux for these excitation temperatures
because
the line is almost certainly saturated 
considering its low flux ratio, 3, to \HthirteenCN(4--3). 
We adopted 3000$\pm$1000 Jy \kms\ as the opacity-corrected HCN(4--3) flux
on the basis of \HthirteenCN(4--3) and \HCfifteenN(4--3) fluxes 
and the abundance ratios
[\twelveC/\thirteenC]$\approx$50 and  
[\fourteenN/\fifteenN]$\approx$300 
\citep[Milky Way inner disk values]{Wilson99}.
We also assumed that the vibrationally excited lines are optically thin, because if they were thick 
they would be as bright as the \vtwo=0 lines.
Another assumption here and in the \Tex\ calculation above is
that all HCN lines arise from the same volume.

The vibrational excitation 
is most likely due to infrared radiation rather than collision for our  \Tvib\ and \Trot. 
This is because collisional excitation would make \Trot\  at \vtwo=1 equal to \Tvib.
If the excitation were due to \HH\ collision at \about200 K the gas needs to be at or above 
the critical density for HCN \vtwo=1, which is
$n_{\rm H_2} \sim 5\times 10^{11}$ \percubiccm\  \citep{Ziurys86}.
Radiative excitation is energetically possible. 
Each rotational state J at the vibrational ground level has associated infrared absorption lines
in the P, Q, and R branches.  
The absorption lines supply energy to the HCN(J,J\minus1) emission line,
and  weaker rotational lines at \vtwo=1,
in the limit of radiation-dominated excitation.
Thus, an energy requirement for IR-pumping is 
\begin{equation}  \label{eq.energy}
	\sum_{i=P,Q,R} f_i \, S_{\rm IR} / \lambda_{\rm IR} 
	\gtrsim
	S_{\rm rot} / \lambda_{\rm rot} ,
\end{equation}
assuming that all the lines have the same velocity width.
Here, $S_{\rm IR}$ is the continuum flux density around the absorption wavelength $\lambda_{\rm IR}$,
$f_i$ is the fractional absorption depth,
and
$S_{\rm rot}$ is the mean flux density of the rotational emission line at the wavelength $\lambda_{\rm rot}$.
Parameters for our case are $\lambda_{\rm IR}$=14 \micron, $S_{\rm 14\mu m}$=2.34 Jy and $f_Q \sim 0.3$
\citep{Lahuis07}, and $\lambda_{\rm rot}$=850 \micron\ and $S_{\rm rot} \lesssim$ 1 Jy for J=4.
The condition of Eq. (\ref{eq.energy}) is satisfied with a wide margin;
the excess energy heats gas through collision.
In this radiative excitation, the observed \Tvib\ indicates that the brightness temperature of the excitation source 
seen from the gas must be $\geq$ 200 K at 14 \micron.

The vibrational excitation significantly affects rotational excitation in the vibrational ground level.
We see this in a model with two vibrational levels, $v$=0 and 1, whose energy gap is $T_0$ in temperature.
Each vibrational level has rotational sub-levels labeled with J.
For a molecule at ($v$,J)=(0,J) there exist approximately $e^{-T_0/T_{\rm vib}}$ molecules at $v$=1 that
will spontaneously decay to the (0,J) state.
Thus the rate of spontaneous transition to (0,J) from $v$=1 is $e^{-T_0/T_{\rm vib}} A_{\rm vib}$,
and radiative excitation of the J level at $v$=0 through $v$=1 needs
\begin{equation}  \label{eq.minTvib}
	e^{-T_0/T_{\rm vib}} A_{\rm vib} \geq A_{\rm rot \, J}
	\; \Leftrightarrow \;
	\Tvib \geq 
	T_0 \! \left/ \ln \frac{A_{\rm vib}}{A_{\rm rot \, J}} \right.,
\end{equation}
where \Avib\ and \Arot\ are the Einstein $A$ coefficients for the vibrational and rotational transitions, respectively.
This is equivalent to Equation (6) of \citet{Carroll81} but is more useful when \Tvib\ is available.
Similarly, the radiative excitation through $v$=1 exceeds collisional (de)excitation among $v$=0 rotational levels
when
\begin{equation}  \label{eq.equivNcol}
	e^{-T_0/T_{\rm vib}} A_{\rm vib} \geq n_{\rm col} \, \gamma _{\rm J, J-1},
\end{equation}
where $n_{\rm col}$ is the number density of the main collision partner and
$\gamma _{\rm J, J-1}$ is the collisional rate coefficient.
In other words, radiative pumping to $v$=1 is equivalent to a
gas density $e^{-T_0/T_{\rm vib}} A_{\rm vib} / \gamma _{\rm J, J-1}$ in terms of
the rotational excitation around J at $v$=0.
Parameters for HCN are 
$T_0$=1024 K and $A_{\rm vib}$=3.7  \persec\  for the bending mode \citep{Ziurys86}
and $\gamma_{\rm J, J-1} \approx 1\times10^{-11}$ \cubiccm\ \persec\  
for HCN--\HH\ collisions at J$\leq$10 and gas kinetic temperatures 30--300 K \citep{Dumouchel10}.
For the observed \Tvib\ of 227 K, 
the rate $e^{-T_0/T_{\rm vib}} A_{\rm vib}$ is 0.04 \persec\ 
and 20 times larger (faster) than $A_{\rm rot\; J=4-3}$.  
The minimum \Tvib\ to satisfy Eq. (\ref{eq.minTvib}) is 86, 137, and 218 K for
J=1, 4, and 10, respectively. 
Thus the observed \Tvib\ satisfies Eq. (\ref{eq.minTvib}) for Js up to about 10.
The IR-pumping through \vtwo=1 matches $n_{\rm H_2} \sim 10^9$ \percubiccm\ for our \Tvib\ 
according to Eq. (\ref{eq.equivNcol}),
although this is somewhat overestimated because collisional transitions are not limited to $|\Delta J|=1$
and half of the vibrational transitions are in the Q branch where $\Delta J=0$.

The relatively low rotational temperature \Trot($v_2$=$1^{1f}$; J\about4)=31$\pm$13 K
probably suggests that the system is not in local thermodynamic equilibrium (LTE).
A caution here is that \Trot\ may be higher and even comparable to \Tvib\ depending on the uncertain 
355 GHz continuum mentioned in \S\ref{s.results}; 
\Trot\ would be $59^{+125}_{-27}$ K if we adopt the lower continuum level
estimated from the LSB. 
A possible reason for the lower rotational temperature than \Tvib\ is that rotational excitation at
low J ($\lesssim 5$) is strongly affected by collision and (sub)millimeter radiation;
note that HCN(4--3) is optically thick. The \Trot\ at \vtwo=1 may be relatively low accordingly.
Rotational temperature can be higher in higher J since collisional excitation in $<$100 K gas
gives way there to IR-pumping as the main excitation mechanism.
This may explain the HCN (rotational) temperature of 300 K ($\pm 30$\%) that
\citet{Lahuis07} estimated from the Q-branch absorption profile under LTE assumption,
because high J population broadens the profile.

For comparison, we analyzed the $v=0$ HCN lines alone
using the large-velocity-gradient model via {\tt RADEX} \citep{vanderTak07, LAMDA05}
and found two types of solutions. 
We used for this the abundance ratios mentioned above
and HCN(1--0) data from \citet{Imanishi04}.
One solution has thermalized HCN at J=4,
$N_{\rm HCN}/\Delta V \! \approx \! 10^{14.7}$ \persquarecm \perkms,
\Tkin $\approx$40 K, and \nHH\ $\geq 10^{7.5}$ \percubiccm.  
Another has subthermal excitation at J=4, moderate opacity of \HthirteenCN, 
and such parameters as $N_{\rm HCN}/\Delta V \! \approx \! 10^{15}$ \persquarecm \perkms,
\Tkin $\geq$ 30 K, and \nHH\ $\approx 10^{6}$ \percubiccm.  
These temperatures are consistent with our \Trot\ at \vtwo=1 and the high column densities
are in accord with the deep obscuration toward the nucleus.
However, these solutions cannot explain the observed HCN \vtwo=1 lines.

To summarize, the vibrationally excited HCN is most likely IR pumped, and
the observed degree of vibrational excitation suggests strong influence of the excitation path through \vtwo=1
to the rotational excitation in the ground state although it may not dominate low-J excitation.
Properties of (dense) molecular gas in NGC 4418 cannot be studied through HCN ignoring the IR-pumping.
In further study, one needs to reexamine our simplistic assumptions that the emitting 
regions of \vtwo=0 and 1 lines are cospatial and have 
a single characteristic temperature and density. 
In reality, gas density must be diverse, and there must be low density gas where HCN 
is excited only under the IR-pumping. 
Thus the total flux of any HCN $v$=0 line should be enhanced at least for that component.
Another improvement in modeling would be to include atomic hydrogen and electrons 
as collision partners \citep{Scoville80,Faure07}.

\section{Other Lines and Other Galaxies}
\begin{figure}[t]
\begin{center}
\epsscale{0.55}
\plotone{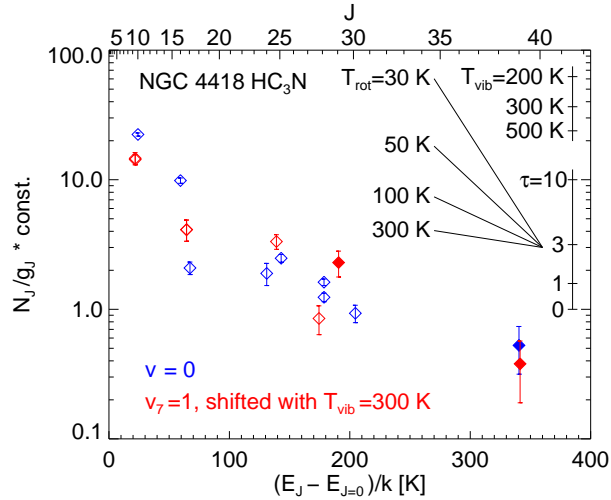} 
\end{center}
\caption{ \label{f.n4418pop}
Population diagram of \HCthreeN.
The lower and upper abscissae are rotational energy measured from the rotational ground level
of each vibrational state and
rotational quantum number $J$, respectively.
The ordinate is population divided by the degree of degeneracy, $N_{J,v,l}/g_{J}$, in arbitrary unit.
Open data points are from \citet{CA10} and filled ones are from our observations.
Error bars are $\pm 1$$\sigma$. 
No correction is made for optical depth, but shifts needed for the correction are shown
in the right side of the plot.
The data for $v_7=1$ lines (red) are shifted against those for the vibrational ground ones (blue)
using a vibrational temperature $\Tvib=300$ K. 
(The $v_7$=1 state is 321 K above $v$=0.)
Shifts for other \Tvib\ are shown in the top-right corner.
To its left are the slopes of rotational population for various rotational temperatures.
}
\end{figure}

\HCthreeN\ is detected in its vibrational ground and excited states though
some transitions are marginal or blended with other lines.
Its population diagram in Fig. \ref{f.n4418pop} shows that the measured line fluxes are
consistent with those of other transitions detected by \citet{CA10}.
The diagram indicates overall excitation temperatures in the range of \about100--300 K, 
except for the $E/k < 100$ K levels where \citet{CA10} estimated \Trot=29 K for $v$=0.  
They also obtained \Tvib \about 500 K.
These temperatures are compatible with our HCN excitation temperatures.

The HCN to \HCOplus\ ratio of the mean brightness temperatures
is $1.65\pm0.07$ for J=4--3 and $1.61\pm0.08$ for J=3--2 
on the assumption of the same emitting volume; it is
1.9 for J=1--0 \citep{Imanishi04}.
Brighter HCN than \HCOplus\ in some Seyfert galaxies and in this galaxy 
has been proposed to be a result of or evidence for chemistry driven by X-ray radiation from 
active nucleus \citep{Kohno01, Imanishi04} although there have been different views
\citep[e.g.,][]{Lahuis07,Baan10}.
It is conceivable at least for objects as deeply embedded as the nucleus of NGC 4418 
that the IR-pumping also affects this ratio.
The minimum brightness temperature of the infrared source needed for 
the radiative pumping of J=1 is given by Eq. (\ref{eq.minTvib}) as
$T_{\rm b}^{\rm\; min}= T_0 / \ln (A_{\rm vib}/A_{\rm rot \, 1-0}) $ 
because 
vibrational temperature is at or below the source brightness temperature 
at the wavelength of the vibrational transition.
This minimum temperature is  86 K at 14 \micron\ for HCN  
and 156, 115,  107, 80, and 33 K 
at 4.6, 7.8, 12, 15, and 45 \micron\ 
for CO, CS, \HCOplus, \NtwoHplus, and \HCthreeN, respectively
\citep[][for $A_{\rm vib}$]{Chandra96,Chandra95,Mauclaire95,Heninger03}.
We note that \HCOplus\ needs higher \Tb\ than HCN for IR-pumping 
and that vibrationally excited \HCOplus(3--2) is not detected at an expected
frequency of 266.8 GHz  
(see Fig. \ref{f.cmpsub_spec}).
IR-pumping may therefore play a role in NGC 4418 
for the high HCN-to-\HCOplus\ ratio
and maybe also for HCN-to-CO ratio and the bright \NtwoHplus.
Its synergy with chemistry in hot galaxy nuclei \citep[e.g.,][]{Harada10} needs 
further study.

We searched published spectra for 
the HCN($\vtwo=1^{1f}$) rotational lines at \plus400 \kms\ of  \HCOplus($v=0$) 
to assess the prevalence of the HCN infrared pumping.
We found no convincing case though a possible sign is at NGC 6240 nucleus in \citet{Wilson08}.
The limitations are sensitivity, spectral baseline, and blending with the brighter \HCOplus\ lines.
The lines must be as narrow as in NGC 4418 for unambiguous detection.

\section{Concluding Remarks}
\label{s.ConcludingRemark}
We evaluated the effect of infrared pumping to the rotational excitation of HCN
using vibrational and rotational temperatures.
The vibrational temperature became measurable with
our detection of rotational lines from vibrationally excited HCN.
The high \Tvib\ in NGC 4418
suggests that HCN is IR pumped and cannot simply trace dense gas in the nucleus.
Other molecules should also be subject to the pumping in different degrees 
according to their energy structures, $A$ coefficients, spatial distributions, 
and the spectrum of the radiation field. 
Observations of \Tvib\ in various molecules and galaxies will tell us the overall significance of 
IR-pumping in extragalactic observations.
Regarding NGC 4418, our high \Tvib\ of HCN agrees with earlier studies to
suggest a $\geq 200$ K excitation source in the nucleus.
The (sub)millimeter lines from vibrationally excited HCN are our new tool to probe hot molecular gas 
close to the nuclear heat source.

\acknowledgements
We thank Daniel Espada for the CO(2--1) spectrum,
Mark Gurwell for advice on flux calibration, 
the SMA operation team for the service observations,
and the referee for helpful comments.
This research extensively used
the NASA/IPAC Extragalactic Database,
NASA's Astrophysics Data System,
the Cologne Database of Molecular Spectroscopy,
and the Splatalogue database.
This work was supported by the grant 99-2112-M-001-011-MY3 from 
the National Science Council of Taiwan.

{\it Facilities:} \facility{SMA}


\clearpage


\begin{deluxetable}{llrcccc}
\tablewidth{0pt}
\tablecaption{Emission Properties of NGC 4418 \label{t.param} }
\tablehead{ 
	\colhead{Emission}  &
	\colhead{\frest} &
	\colhead{$E_{\rm u}/k$}  &	
	\colhead{$F({\rm line}), S_\lambda$}  &	
	\colhead{$\sigma_{\rm th} $} &
	\colhead{$\sigma_{\rm tot} $} &				
	\colhead{note}
	\\
	\colhead{ }  &
	\colhead{[GHz]} &
	\colhead{[K]}  &	
	\multicolumn{3}{c}{[Jy \kms, mJy]}  &
	\colhead{ }
	\\
	\colhead{(1)}  &	
         \colhead{(2)}  &	
         \colhead{(3)}  &
	\colhead{(4)}  &	
	\colhead{(5)}  &	
	\colhead{(6)}  &
	\colhead{(7)}	
}
\startdata
\HCOplus(J=4--3)      & 356.734 &42.8& 82 & 3 & 9 &      \\
HCN\vJfourtothree    & 356.256 &1067.1& 30 & 3 & 4 &  \\
\HCthreeN($v_7$=$1^{\rm 1e}$,J=39--38) & 355.566 &662.2& 6 & 3 & 3 & (a) \\
\HCthreeN(J=39--38) & 354.697 & 340.5 & 25 & \nd & 10 & (b) \\
HCN(J=4--3)               & 354.505 &42.5& 133 & 3 &14 & (c)  \\
CO(J=3--2)                  & 345.786 &33.2& 746 & 3 & 75 & (d) \\
CO(J=3--2), 10\asec  & 345.796 &33.2& 782 & 3 & 78 & (d) \\
\HthirteenCN(J=4--3) & 345.340 &41.4& 41 & 4 & 5  & (e) \\
\HCfifteenN(J=4--3)   & 344.200 &41.3& 11 & 3  & 3 &   \\
CS(J=7--6)                  & 342.883 &65.8& 63 & 3  & 7 &      \\
\NtwoHplus(J=3--2)   & 279.512 &26.8& 25 & 2  & 3 &      \\
\HCOplus(J=3--2)       & 267.558 &25.7& 59 & 2 & 6 &      \\
HCN\vJthreetotwo      & 267.199 &1050.0& 16 & 2 &  3 &     \\
HCN(J=3--2)               & 265.886 &25.5& 94 & 2 & 10  &      \\
\HCthreeN($v_7$=$1^{\rm 1f}$,J=29--28) & 264.817 &511.5& 11 & 2 & 3 & (f) \\
0.85 mm continuum   &\nd&\nd& 210 & 5 & 22 & (g) \\
0.88 mm continuum   &\nd&\nd& 169 & 3 & 17 & (g) \\
1.08 mm continuum   &\nd&\nd&  96 &  2 & 10 & (g) \\
1.13 mm continuum   &\nd&\nd&  86 & 3  &   9  & (h)
\enddata
\tablecomments{
(2) Line frequency.
(3) Upper level energy.  
(4) Line flux in  Jy \kms\  and continuum flux density in mJy
in the central 6\arcsec\ (1 kpc).
Line fluxes are integrated over 1900--2300 \kms\
unless otherwise noted and are continuum subtracted.
No correction is made for missing flux.
CO(3--2) flux is also given for the central 10\asec. 
No other emission was resolved by the 6\asec\ beam.
(5) Thermal error of (4).
For lines, this includes thermal noise of continuum propagated through
continuum subtraction. 
(6) Total error of (4) including $\sigma_{\rm th}$ and 10\% error of flux calibration.
(a) Marginal detection with possible blending with weaker lines.
(b) Large uncertainty due to line blending.
(c) No subtraction is made for the blended \HCthreeN(39--38).
(d) There is likely contamination by \HCthreeN(J=38--37) at \frest=345.609  GHz \citep{CA10}.
(e) This line is blended with CO(3--2) and likely with \HCthreeN(38--37).
Its flux is obtained by doubling the line flux
in 2100--2300 \kms\ assuming a symmetric profile.
(f) Flux from 1900--2197 \kms.
(g) Continuum flux densities are measured outside the 1900--2300 \kms\ velocity ranges of
the lines marked in Fig. \ref{f.cmpsub_spec}. Also excluded for 0.85 mm continuum are
HCN(4--3) velocities of 2300--2320 \kms.
(h) Flux density is measured at the \HCOplus(3--2) velocities below 1900 \kms.
}
\end{deluxetable}

\clearpage

\end{document}